\documentclass[aps,prd,amsfonts,groupedaddress,showpacs,showkeys]{revtex4}
\usepackage{epsfig}
\newcommand{\eq}{\begin{eqnarray}}
\newcommand{\en}{\end{eqnarray}}
\newcommand{\bea}{\begin{eqnarray}}
\newcommand{\eea}{\end{eqnarray}}

\def\d{D^0}
\def\db{\bar{D}^{0}}
\def\ds{D^{\ast \, 0}}
\def\dbs{\bar{D}^{\ast \, 0}}

\begin{document}

\title{
{\textbf{$X(3872)$ as a molecular $DD^*$ state in a potential model}}}
\author{
Ian Woo Lee\footnote{E-mail: ian-woo.lee@student.uni-tuebingen.de},
Amand Faessler,
Thomas Gutsche,
Valery E. Lyubovitskij\footnote{On leave of absence
        from Department of Physics, Tomsk State University,
        634050 Tomsk, Russia}
\vspace*{1.2\baselineskip}}

\affiliation{Institut f\"ur Theoretische Physik, Universit\"at T\"ubingen, \\
Kepler Center for Astro and Particle Physics, \\
Auf der Morgenstelle 14, D-72076 T\"ubingen, Germany
\vspace*{0.3\baselineskip}\\}

\date{\today}

\begin{abstract}

We discuss the possibility that the $X(3872)$ can be a hadronic $DD^*$
bound state in the framework of a potential model. The potential is
generated by the exchange of pseudoscalar, scalar and vector mesons
resulting from the Lagrangian of heavy hadron chiral perturbation theory.
The hadronic bound state configuration contains charged and neutral
$DD^*$ components, while orbital $S$- and $D$-waves are included.
Isospin symmetry breaking effects are fully taken into account.
We show that binding in the $DD^*$ system with $J^{PC}=1^{++}$
already exists for a reasonable value of the meson-exchange
regularization parameter of $\Lambda \sim 1.2$ GeV. We also explore
the possibility of hadronic $BB^*$ bound states and show that binding
can be achieved in the isoscalar limit for $J^{PC}=1^{++}$ or $1^{+-}$.

\end{abstract}



\maketitle

\section{Introduction}

The new resonance $X(3872)$ has been discovered a few years ago by
Belle~\cite{Choi:2003ue} and has been later
confirmed by the CDF~\cite{Acosta:2003zx}, D0~\cite{Abazov:2004kp}
and {\it BABAR}~\cite{Aubert:2004ns} collaborations.
It is classified as an isosinglet state with positive charge
parity. Current averaged results for the $X(3872)$ mass and width are:
$m_X=3872.2 \pm 0.8$ MeV and
$\Gamma_X = 3.0^{+1.9}_{-1.4} \pm 0.9$~MeV~\cite{Amsler:2008zzb}.
For the $X(3872)$ several structure interpretations have been
proposed in the literature (for a status report see e.g.
Refs.~\cite{Swanson:2006st,Bauer:2005yu,Voloshin:2007dx}).
In the context of molecular
approaches~\cite{Voloshin:1976ap}-\cite{Liu:2008tn} the $X(3872)$
can be identified with a weakly--bound hadronic molecule whose constituents
are $D$ and $D^\ast$ mesons. The reason for this natural interpretation is
that $m_X$ is very close to the $\d \dbs$ threshold and hence is in analogy
to the deuteron --- a weakly--bound state of proton and neutron.
Note, that the idea to treat charmonium-like states as hadronic molecules
traces back to Refs.~\cite{Voloshin:1976ap,DeRujula:1976qd}.
Originally it was proposed that the state $X(3872)$ is a superposition
of $\d \dbs$ and $\db \ds$ pairs. Later (see e.g. discussion
in Refs.~\cite{Swanson:2003tb,Voloshin:2004mh,Braaten:2005ai}) also
other structures, such as a charmonium state or even other
meson pair configurations, were discussed in addition to the
$\d  \dbs +$ charge conjugate (c.c.) component.
The possibility of two nearly degenerated $X(3872)$ states with positive
and negative charge parity has been discussed in
Refs.~\cite{Terasaki:2007uv,Gamermann:2007fi}.

In several papers~\cite{Tornqvist:2004qy,
Liu:2008fh,Thomas:2008ja,Liu:2008tn} the possibility that the
$X(3872)$ can be interpreted as a hadronic molecule ---
a bound state of $\d$ and $\ds$ mesons
--- has been investigated in potential models.
These approaches can be traced back to
Ref.~\cite{Tornqvist:1993ng}, where possible
deuteron--like two--meson bound states have been considered in the context
of a potential generated by the pion exchange mechanism.
It was shown that
bound states of two mesons are possible in analogy to a deuteron model
with one--pion exchange only. In addition, the relevance of the tensor
interaction has been pointed out~\cite{Tornqvist:1993ng} which connects
different spin-orbit configurations.
These tensorial terms were found to be important to generate bound states
in the potential model of the deuteron with a reasonable value for the
cutoff $\Lambda$, which is a parameter regularizing the hadronic interaction.
The value of $\Lambda$ is determined phenomenologically and depends on the
model, though its scale in low--energy hadron physics is of the order of 1 GeV.
Later, after the discovery of the $X(3872)$,
Tornqvist also showed~\cite{Tornqvist:2004qy}
the importance of isospin breaking effects in a possible binding mechanism
because of the mass differences between the neutral and charged
$D (D^\ast )$ mesons. Thomas and Close reinvestigated~\cite{Thomas:2008ja}
in the context of one-pion exchange
if the $X(3872)$ can be a $J^P =1^+$ $D\bar D^\ast$ bound state including
charged mode and $D$-wave configurations. They also point out technical
differences with respect to previous approaches, but moderate binding
depends rather sensitively on the parameters involved.
However, Liu {\it et al.}~\cite{Liu:2008fh} claim that when
taking into account both pion and also sigma meson exchange
potentials the $D$ and $D^*$
mesons cannot form a bound state with reasonable values for the potential
parameters.
They show that a bound state is not present for values of $\Lambda < 5.8$ GeV,
where the upper value is too large compared to the typical hadronic scale of
$\Lambda \simeq 1$ GeV. But in their calculation only the $S$-wave
configuration was taken into account.
Moreover, only the interaction transitions
$D^0\bar{D}^{*0} \rightarrow D^{*0}\bar{D}^0$ involving neutral $D$ mesons
were considered. This causes that the interaction strength is reduced by
a factor of $1/3$ compared to that of the isoscalar state in the isospin
symmetry limit. Later, Liu {\it et al.}~\cite{Liu:2008tn} showed
that the $X(3872)$ is open to the possibility of a loose bound state of
$D\bar{D}^*$ when one considers the isospin symmetry limit with $I=0$ and
also when further heavy meson-exchange, such as
$\rho$ or $\omega$ exchange, are included.
Again, they restricted the calculation to the $S$-wave component
and to the isospin symmetry limit.

The main objective of the present work is to improve the description of the
$X(3872)$ as a possible bound state of $D$ and $D^\ast$ mesons in the
context of
the potential model considered previously in Refs.~\cite{Tornqvist:1993ng,%
Tornqvist:2004qy,Thomas:2008ja,Liu:2008fh,Liu:2008tn}.
In particular, we first consider a full set of
mesons ($\pi$, $\sigma$, $\eta$, $\rho$ and $\omega$) in constructing the
nonrelativisitic one--meson exchange potential, then we include charged
and neutral $D \bar D^\ast$ components both in orbital $S$- and $D$-waves.
Full account is taken of isospin breaking effects by incorporating mass
differences in the D and $D^\ast$ mesons.
As in the preceding papers we also regularize the
potential by introducing form factors containing the cutoff parameter
$\Lambda$.

In the manuscript we proceed as follows. First, in Sec.~II
we discuss the basic notions of our approach. We define the effective
meson Lagrangian based on heavy hadron chiral perturbation theory,
which then is used for the derivation of the meson-exchange
potential in coordinate space. The formalism is developed to explicitly
consider S--D wave mixing and also to include isospin--symmetry breaking
effects.
In Sec.~III we first discuss the numerical procedure for solving the
coupled-channel Schroedinger equation. Then we present our results for
a possible binding of the $D \bar D^\ast$ system, discussing the influence
of various components both in the potential and the di-meson configuration.
In Section IV we extend our approach to the $B B^\ast$ bound state system
as well. Finally, in Sec.~V, we present a short summary of our results and
some general comments.
\section{Method}
\subsection{Effective Lagrangian}
Our starting point (as in the potential approach pursued in
Refs.~\cite{Liu:2008fh,Liu:2008tn})
is the effective Lagrangian of heavy hadron chiral perturbation theory
(HHChPT)~\cite{Wise:1992hn,Isola:2003fh} based on chiral and heavy quark
symmetries. It gives rise to the interaction Lagrangians
between $D(D^\ast)$ mesons and light pseudoscalar, scalar and vector
mesons with
\eq
{\cal L}_{{\cal D}{\cal D}^*\mathbb{P}}&=&
-ig_{{\cal D}{\cal D}^*\mathbb{P}}({\cal D}_a{\cal D}_{\mu b}^{* \dagger}
-{\cal D}_{\mu a}^*{\cal D}_b^\dagger)\partial^\mu\mathbb{P}_{ab}\,,
\nonumber\\
{\cal L}_{{\cal D}{\cal D}^*\mathbb{V}}&=&-2f_{{\cal D}{\cal D}^*\mathbb{V}}
\varepsilon_{\mu\nu\alpha\beta}(\partial^\mu\mathbb{V}^\nu)_{ab}
[({\cal D}_a^\dagger\partial^\alpha{\cal D}_b^{*\beta}
-\partial^\alpha{\cal D}_a^\dagger{\cal D}_b^{*\beta})
-({\cal D}_a^{*\beta\dagger}\partial^\alpha{\cal D}_b
-\partial^\alpha{\cal D}_a^{*\beta\dagger}{\cal D}_b)] \,, \nonumber\\
{\cal L}_{{\cal DD}\sigma}&=&-2m_{\cal{D}}
g_\sigma{\cal D}_a {\cal D}_a ^\dagger \sigma \,, \\
{\cal L}_{{\cal D}^*{\cal D}^*\sigma}&=&2m_{{\cal D}^*}
g_\sigma{\cal D}_a^{*\alpha}{\cal D}_{\alpha a}^{*\dagger}\sigma \,,
\nonumber\\
{\cal L}_{{\cal DD}\mathbb{V}}&=&-ig_{{\cal DD}\mathbb{V}}({\cal D}_a^\dagger
\partial_\mu {\cal D}_b-{\cal D}_b \partial_\mu
{\cal D}_a^\dagger)(\mathbb{V}^\mu)_{ab}\,,
\nonumber\\
{\cal L}_{\cal{D}^*\cal{D}^*\mathbb{V}}&=&ig_{\cal{D}^*\cal{D}^*\mathbb{V}}
({\cal D}_a^{*\nu\dagger}\partial^\mu{\cal D}_{\nu,b}^*-
{\cal D}_{\nu,b}^*\partial^\mu{\cal D}_a^{*\nu\dagger})(\mathbb{V}_\mu)_{ab}
+4if_{\cal{D}^*\cal{D}^*\mathbb{V}}{\cal D}_{\mu, a}^{*\dagger}
{\cal D}_{\nu, b}^*(\partial^\mu\mathbb{V}^\nu
-\partial^\nu \mathbb{V}^\mu)_{ab}
\,,\nonumber
\en
where ${\cal D}^{(*)}=({D}^{0(*)},D^{+(*)},D_s^{+(*)})$. The octet
pseudoscalar $\mathbb{P}$ and the nonet vector $\mathbb{V}$
meson matrices are defined as
\eq\label{eq:pseudo}
\mathbb{P}=\left(
\begin{array}{ccc}\frac{\pi^0}{\sqrt{2}}+\frac{\eta}{\sqrt{6}}&\pi^+&K^+\\
\pi^-&-\frac{\pi^0}{\sqrt{2}}+\frac{\eta}{\sqrt{6}}&K^0\\
K^-&\bar{K}^0&-\frac{2\eta}{\sqrt{6}}\\
\end{array} \right)
\en
and
\eq\label{eq:vector}
\mathbb{V}=\left( \begin{array}{ccc}\frac{\rho^0}{\sqrt{2}}
+\frac{\omega}{\sqrt{2}}&\pi^+&K^{*+}\\
\rho^-&-\frac{\rho^0}{\sqrt{2}}+\frac{\omega}{\sqrt{2}}&K^{*0}\\
K^{*-}&\bar{K}^{*0}&\phi\\
\end{array} \right) \,.
\en
The HHChPT couplings
are given as
\eq\label{eq:HHChPT_par}
& &g_{{\cal D}{\cal D}^*\mathbb{P}}=
\frac{2g}{f_\pi}\sqrt{m_{\cal D}m_{\cal D^*}}\,, \hspace*{.25cm}
g_{{\cal D}{\cal D}\mathbb{V}}=g_{{\cal D}^*{\cal D}^*\mathbb{V}}
=\frac{\beta g_{\mathbb{V}}}{\sqrt{2}} \,,\nonumber\\
& &f_{{\cal D}{\cal D}^*\mathbb{V}}
=\frac{f_{{\cal D}^*{\cal D}^*\mathbb{V}}}{m_{\cal D^*}}
=\frac{\lambda g_{\mathbb{V}}}{\sqrt{2}}\,, \hspace*{.25cm}
g_{\mathbb{V}}=\frac{m_\rho}{f_\pi}\,, \hspace*{.25cm}
g_\sigma=\frac{g_\pi}{2\sqrt{6}} \,, \\
& &g=0.59\,, \hspace*{.25cm}
\beta=0.9\,, \hspace*{.25cm}
\lambda=0.56 \ {\rm GeV}^{-1}\,,
\hspace*{.25cm}
f_\pi=132 \ {\rm MeV}\,,
\hspace*{.25cm}
g_\pi=3.73\,. \nonumber
\en

We assume that the $X(3872)$ is a bound state of $D^*$and $D$ with
quantum numbers $J^{\rm PC}=1^{++}$. To compose a $1^{++}$ state with
a pseudoscalar and a vector meson, the orbital angular momentum can have
the values of $L=0$ and $2$.
The leading--order meson exchange diagrams constructed with the use of
these Lagrangians are shown in Figs.~1 (pseudoscalar-vector direct channels)
and 2 (pseudoscalar-vector crossed channels).
All diagrams are fixed to conserve angular momentum and parity.
In the isospin symmetry limit, where $D^{(*)0}$ and $D^{(*)\pm}$ have no mass
difference, the $X(3872)$ can be represented explicitly as
\begin{equation}
\label{eq:X_state}
|X\rangle=\frac{1}{2}[(|D^0\bar{D}^{*0}\rangle-c|D^{*0}\bar{D}^0\rangle)
\pm(|D^+D^{*-}\rangle-c|D^{*+}D^-\rangle)] \end{equation}
where the $\pm$ sign is determined according to the total isospin.
The $+$ sign corresponds to the isovector state and the $-$ sign to
the isoscalar one;
the sign $c=\pm1$ adjusts even and odd charge conjugation parity, respectively.
Here we use the convention where the vector field changes sign
after charge conjugation.
(for a detailed discussion about this issue and conventions used in
literature see Refs.~\cite{Liu:2008fh,Thomas:2008ja}).

\subsection{Effective potentials in the direct channels}
The effective potentials in momentum space are derived for
the direct channels as follows :
\eq\label{V_pot_sig}
V_{\rm dir}(q)&=&g_{\sigma}^2\frac{1}{q^2-m_\sigma^2}
-\frac{\gamma}{2}g_{{\cal D}{\cal D}\mathbb{V}}
g_{{\cal D}^*{\cal D}^*\mathbb{V}}
\left[\frac{1}{q^2-m_{\rho}^2}+\frac{{q}^2}{4m_Dm_{D^*}m_\rho^2}\right]
\nonumber\\
&+&\frac{1}{2}g_{{\cal D}{\cal D}\mathbb{V}}g_{{\cal D}^*{\cal D}^*\mathbb{V}}
\left[\frac{1}{q^2-m_{\omega}^2}+\frac{{q}^2}{4m_Dm_{D^*}m_\omega^2}\right]
\en
where $\gamma=1$ and $-3$ corresponds to the isovector and
isoscalar channel, respectively.
To avoid
the singular behavior at small distances we further regularize the potential
with a form factor
$F(q^2)=\frac{\Lambda^2-m^2}{\Lambda^2-q^2}$ ($m$ is the mass of the
exchanged meson) at each vertex.
The value of the parameter $\Lambda$ is not strictly determined.
It should be about $1$ GeV (a typical scale in low--energy
physics) though its specific value can depend on the particular application.
A larger value of $\Lambda$ enhances the potential at short distances
hence possible binding energies will depend on $\Lambda$
as we shall discuss in Sec.~IV.

The potentials in coordinate space can be obtained by Fourier
transformation :
\begin{eqnarray}
\label{eq:V_dir}
V_{\rm dir}(r)&=&
-\kappa_1 V_0 P_\sigma (r)
+\gamma [\kappa_2 V_0P_\rho (r) +\kappa_3 V_0Q_\rho (r)]
-[\kappa_2 V_0P_\omega (r) +\kappa_3 V_0Q_\omega (r)]
\end{eqnarray}
with the radial dependence
\begin{equation}
P_M(r)=\frac{m_\rho^2}{m_\pi^2}\Big(\frac{e^{-\mu_M r}-e^{-\chi r}}{m_\pi r}
-\frac{\Lambda^2-m_M^2}{2m_\pi\chi}e^{-\chi r}\Big)\,,
\end{equation}
\begin{eqnarray}
\label{eq:Q(r)}
Q_M(r)=\frac{1}{m_\pi^3m_M^2}
(\Lambda^2-m_M^2)^2\left(1-\frac{\Lambda^2}{2\chi}r \right)
\frac{e^{-\chi r}}{r} \,,
\end{eqnarray}
where $\mu_M^2=m_M^2-(m_{1'}-m_1)^2$ and $\chi^2=\Lambda^2-(m_{1'}-m_1)^2$.
The subscripts at $P$ and $Q$ in Eq.~(\ref{eq:V_dir}) refer to
the corresponding exchanged meson of mass $m_M$.
Here we also define the constant $V_0$
\eq
V_0\equiv\frac{m_\pi^3}{12\pi}\frac{g^2}{f_\pi^2},
\en
which was already used in
Refs.~\cite{Tornqvist:1993ng,Thomas:2008ja}.
This quantity is introduced to compare the strength of the one--pion
exchange potential to the counterparts from the exchange of other scalar and
vector mesons. The relevant strength coefficients are listed in the following.
They are all dimensionless and the dimension of the potential is
carried by $V_0$:
\begin{eqnarray}
\kappa_1&=&\frac{1}{8}\frac{g_\pi^2}{g^2}\frac{f_\pi^2}{m_\rho^2}\simeq0.146,\\
\kappa_2&=&\frac{3}{4}\frac{\beta^2}{g^2}\simeq1.75,\\
\kappa_3&=&\frac{3}{16}
\frac{\beta^2}{g^2}\frac{m_\rho^2}{m_Dm_{D^*}}\simeq0.070.
\end{eqnarray}

\subsection{Effective potentials in the crossed channels}

In the crossed channels we get the following effective potential
in momentum space:
\begin{eqnarray}
\label{eq:crosslagrangian}
V_{\rm cross}(q)&=&
-\gamma c\frac{g^2_{{\cal D}{\cal D}^*\mathbb{P}}}
{24m_{{\cal D}}m_{{\cal D}^*}}
\frac{3(\vec{\varepsilon_1}\cdot\vec{q})
(\vec{\varepsilon^*_{2'}}\cdot\vec{q})}{q^2-m^2_{\pi}}
+c\frac{g^2_{{\cal D}{\cal D}^*\mathbb{P}}}{72m_{{\cal D}}m_{{\cal D}^*}}
\frac{3(\vec{\varepsilon_1}\cdot\vec{q})(\vec{\varepsilon^*_{2'}}
\cdot\vec{q})}{q^2-m^2_{\eta}}
\nonumber\\
&+&2\gamma cf^2_{{\cal D}{\cal D}^*\mathbb{V}}
\frac{\vec{q}^2(\vec{\varepsilon}_1\cdot\vec{\varepsilon}_{2'}^*)-(\vec{\varepsilon_1}
\cdot\vec{q})(\vec{\varepsilon^*_{2'}}\cdot\vec{q})}{q^2-m^2_{\rho}}
-2cf^2_{{\cal D}{\cal D}^*\mathbb{V}}
\frac{\vec{q}^2(\vec{\varepsilon}_1\cdot\vec{\varepsilon}_{2'}^*)-(\vec{\varepsilon_1}
\cdot\vec{q})(\vec{\varepsilon^*_{2'}}\cdot\vec{q})}{q^2-m^2_{\omega}}
\end{eqnarray}
where $\vec{\varepsilon}_1$ and $\vec{\varepsilon}_{2'}^*$
are the polarization vectors of ingoing and outgoing vector mesons,
respectively.

The quantity $c=\pm1$, related to charge conjugation in
Eq.~(\ref{eq:X_state}), appears in every crossed diagram,
which is different from the direct channels.
The crossed channel potentials in coordinate space are obtained
in similar manner as in the previous case:
\eq
V_{\rm cross}(r)&=&c\Big\{-\frac{\gamma}{2}V_0[C^-_{\pi}(r)+S_{12}T^-_{\pi}(r)]
+\frac{1}{6}V_0[C^+_{\eta}(r)+S_{12}T^+_{\eta}(r)]
\nonumber \\
&+&\kappa_4 \gamma V_0[C^+_{\rho}(r)-\frac{1}{2}S_{12}T^+_{\rho}(r)]
-\kappa_4 V_0[C^+_{\omega}(r)-\frac{1}{2}S_{12}T^+_{\omega}(r)]\Big\},
\en
where
\eq
\label{eq:Cm}
C_{}^+(r)&=&\frac{\mu^2}{{m_\pi}^2}\frac{e^{-\mu r}-e^{-\chi r}}{m_\pi r}
-\frac{\chi(\Lambda^2-{m_{}}^2)}{2{m_\pi}^3}e^{-\chi r},\\
\label{eq:Tm}
T^+(r)&=&(\mu^2r^2+3\mu r+3)\frac{e^{-\mu r}}{{m_\pi}^3 r^3}
-(\chi^2r^2+3\chi r+3)\frac{e^{-\chi r}}{{m_\pi}^3r^3}
-\frac{\Lambda^2-{m_{}}^2}{{m_\pi}^2}(\chi r+1)\frac{e^{-\chi r}}{2{m_\pi}r},\\
S_{12}&=&3(\vec{\varepsilon}_{1}\cdot\hat{r})
(\vec{\varepsilon}_{2'}^{\, \ast}\cdot\hat{r})
-(\vec{\varepsilon}_{1}\cdot\vec{\varepsilon}_{2'}^{\, \ast}).
\en
Here $m$ is the mass of the exchanged meson; $\mu$ and $\chi$ are defined as
$\mu^2=m^2-(m_{D^*}-m_D)^2$ and $\chi^2=\Lambda^2-(m_{D^*}-m_D)^2$,
respectively;
$C^+$ and $T^+$ are the potentials for positive values of $\mu^2$.
In the case of $\pi$ exchange, especially for $\pi^0$ exchange,
the mass of $\pi^0$ is smaller than the mass difference
$(m_{D^*}-m_D)$ so that $\mu^2$ becomes negative.
To take care of this case we define $\bar{\mu}^2 = -\mu^2$ and obtain the real
parts of these potentials as
\eq
C^-(r)&=&-\frac{\bar{\mu}^2}{{m_\pi}^2}\frac{cos(\bar{\mu} r)
-e^{-\chi r}}{m_\pi r}
-\frac{\chi(\Lambda^2-{m}^2)}{2{m_\pi}^3}e^{-\chi r},\\
\label{eq:g}T^-(r)&=&(-\bar{\mu}^2r^2+3)\frac{cos(\bar{\mu} r)}{{m_\pi}^3r^3}
+3\bar{\mu} r\frac{sin(\bar{\mu} r)}{{m_\pi}^3r^3}
-(\chi^2r^2+3\chi r+3)\frac{e^{-\chi r}}{{m_\pi}^3r^3}
-\frac{\Lambda^2-{m}^2}{{m_\pi}^2}(\chi r+1)\frac{e^{-\chi r}}{2m_\pi r}.
\end{eqnarray}
$\Lambda$ is sufficiently large so
that the sign of $\chi^2$ is not affected by the sign of $\mu^2$.

The functions $C^\pm$ and $T^\pm$ are dimensionless and their shape depends
on the value of $\Lambda$ and the mass $m$ of the exchanged mesons.
To compare the strength of the potential due to vector meson exchange
to the counterparts of the pseudoscalar mesons we define
the dimensionless parameter $\kappa_4$ as
\eq
\kappa_4=\frac{4}{3}f^2_{{\cal D}^*{\cal D}\mathbb{V}}\frac{3f_\pi^2}{g^2}
=2\lambda^2\frac{m_\rho^2}{g^2}
\simeq 1.07.
\en

We present the potentials in the crossed channels again
to see the tensorial terms clearly. Writing the potential in the
$L=0$, $2$ basis we have
\begin{equation}
V_{\rm cross}(r)= V_0\left[ \left(\begin{array}{cc}1&0\\0&1\end{array}
\right)C_{\rm cross}(r)
+\left( \begin{array}{cc}0&-\sqrt{2}\\-\sqrt{2}&1\end{array}\right)
T_{\rm cross} (r) \right]
\end{equation}
where
\eq
C_{\rm cross}(r)&=&-\frac{\gamma}{2}C_\pi^-(r)+\frac{1}{6}C_\eta^+(r)
+\gamma\kappa_4C_\rho^+(r)-\kappa_4C_\omega^+(r),\\
T_{\rm cross}(r)&=&-\frac{\gamma}{2}T_\pi^-(r)+\frac{1}{6}T_\eta^+(r)
-\frac{\gamma}{2}\kappa_4T_\rho^+(r)+\frac{1}{2}\kappa_4T_\omega^+(r)~.
\en

Finally we get the total meson exchange potential, which is the sum of
the potentials from the direct and crossed channels:
\eq
V_{\rm total}(r) = V_{\rm dir} (r) + V_{\rm cross} (r).
\en

The profile functions $P(r)$, $Q(r)$, $C(r)$ and $T(r)$ encoding the
contribution of meson exchange to the potential are plotted in Fig.3.
One can see that the functions $C(r)$ and $Q(r)$ dominate at small
distances.
Of course strength and range of all curves depend on the mass of the exchanged
meson and the value of $\Lambda$, but Fig. 3 contains at least the
characteristics of each function. $V_0$ is fixed at 1.3 MeV (or 1.5 MeV)
by the experimental data on the decay
width $\Gamma(D^{*+} \rightarrow D^0\pi^+)$.
We choose the value $V_0$=1.3 MeV used before
in Refs.~\cite{Tornqvist:1993ng,Thomas:2008ja} which is also consistent
with the parameters of the HHChPT Lagrangian~(\ref{eq:HHChPT_par}).

Multiplying the relevant coefficients such as $\kappa_1V_0$,
$\kappa_2 V_0$, etc. with $P(r)$,
$Q(r)$, $C(r)$ and $T(r)$ in Fig.~4 we indicate the potentials in the
isoscalar limit (that is, with $\gamma=-3$).
The total potential is written as
\begin{eqnarray}
V_{total}&=&V_C\left(\begin{array}{cc}1&0\\0&1\end{array}\right)
+V_T\left(\begin{array}{cc}0&-\sqrt{2}\\-\sqrt{2}&1\end{array}\right)
\end{eqnarray}
in the $L=0$, $2$ basis.
The tensor potential $V_T=V_0T_{\rm cross}$ arises only from the
crossed channels. The central potential
$V_C=V_{\rm dir} + V_0C_{\rm cross}$
does not mix $S$ and $D$ waves.
In Fig.4(a) we indicate the individual contributions to $V_C$ and the
total result for a cutoff in the form factors of $\Lambda=1250$ MeV.
The contributions of $\pi$ and $\rho$ meson exchange are dominant and
attractive. The effect of $\sigma$--exchange is small, though it depends
on the $\sigma$ mass.
The vector meson potentials become slightly attractive around 0.1 - 0.2 fm
but are negligible compared to the total potential.
In Fig.4(b) we indicate the off--diagonal potential $V_T$ including all
exchanged mesons for $\Lambda$=1250 MeV.
The pseudoscalar and vector meson potentials have different sign, but the
$\pi$ contribution dominates such that the total potential becomes attractive.
>From Eq.~(\ref{eq:crosslagrangian}) it is evident that the different signs
of the pseudoscalar and vector meson potentials are
dictated by the opposite sign in the effective Lagrangian.
Not only $V_C$ plays a possible role to generate attraction for the
X state but also $V_T$ though its strength is small compared to $V_C$ at short
distances.

\subsection{Isospin symmetry breaking}
We set up the $X(3872)$ as a mixture of the isoscalar and the isovector
components,
that is
$|X\rangle = c_0 |0 \rangle + c_\pm|\pm \rangle$, where
$|0\rangle = \frac{1}{\sqrt{2}}(|D^0\bar{D}^{*0}\rangle
-c|D^{*0}\bar{D}^0\rangle)$ and
$|\pm\rangle = \frac{1}{\sqrt{2}}(|D^+\bar{D}^{*-}\rangle
-c|D^{*+}\bar{D}^-\rangle)$.
In the limit of isospin symmetry the coefficients exactly fulfill
the relation
$|c_0|=|c_\pm|=\frac{1}{\sqrt{2}}$. The mass differences between the
neutral and charged channel induce slight changes in these
coefficients. When isospin symmetry is broken, the isoscalar factor
$\gamma=-3$ in the potentials
should be replaced by a $2 \times 2$ matrix
\eq
\left(\begin{array}{rr}-1&-2\\-2&-1\end{array}\right)
\en
in the particle basis of neutral $|0\rangle$ and charged $|\pm\rangle$
states. Here the diagonal matrix
elements of $-1$ come from the exchange of neutral
mesons, while the off-diagonal matrix elements $-2$ arise from charged
meson exchange.
We have four different potentials according to the corresponding diagrams.
Then the Schr\"odinger equation becomes as follows:
\eq
\label{eq:isobreaking}
\left[\left(\begin{array}{cc}M_0
-\frac{\nabla^2}{2m_0}&0\\0&M_\pm-\frac{\nabla^2}{2m_\pm}\end{array}\right)
+\left(\begin{array}{cc}-V_a & -2V_c\\-2V_d & -V_b\end{array}\right)\right]
\left(\begin{array}{c}c_0|0\rangle \\ c_\pm|\pm\rangle\end{array}\right)
=
E\left(\begin{array}{c}c_0|0\rangle \\ c_\pm|\pm\rangle\end{array}\right)~,
\en
where $m_0$ and $m_\pm$ are the reduced masses while
$M_0$, $M_\pm$ are the total masses of neutral and charged systems,
respectively. Here, for example, $V_a$ is the potential which is generated by
the diagrams (a) in Figs.1 and 2 with
\eq
V_a=
\left(\begin{array}{cc}1&0\\0&1\end{array} \right)V_a^{\rm dir}
+\left[ \left(\begin{array}{cc}1&0\\0&1\end{array} \right)
V_0 C_a^{\rm cross}
+\left( \begin{array}{cc}0&-\sqrt{2}\\
-\sqrt{2}&1\end{array}\right)V_0 T_a^{\rm cross} \right]
\en
in the $L=0$, $2$ basis.
Accordingly, $V_{b,c,d}$ are due to the corresponding graphs of  Figs.1 and 2.

\section{Results}
\subsection{Numerical method}
To get solutions to the Schr\"odinger equation with the derived potential
we need to solve coupled second--order
differential equations. The numerical procedure is as follows:
we discretize $r$-space
and diagonalize the potential at each discretized position
with the boundary conditions that both $S$ and $D$ waves vanish
at $r \rightarrow \infty$.
Then the Hamiltonian becomes a finite matrix and we solve
the Schr\"odinger equation using diagonalization routines.
This method is well suited for obtaining the ground state wave function
and eigenvalue in the bound system. We tested the results varying
the  values of the boundary position and the number of discretization points.

To confirm our numerical calculations we also adopt another method
in solving the Schr\"odinger equation by using MATSCS, which is a Matlab
package implementing the CPM\{P,N\} methods for the numerical solution
of the multichannel Schr\"odinger eigenvalue problem~\cite{Ledoux:2007ld}.
We obtain agreement in the results for the energy eigenvalue within 0.1 MeV
between our matrix method
and MATSCS. This double check can only be performed in the limit of
isospin symmetry, because MATSCS can only solve symmetric potentials.
When isospin symmetry is broken the multichannel potentials are slightly
distorted and do not coincide any more for the neutral and the charged states.
\subsection{Results}

In Table~\ref{table:isobreaking} we summarize our results for the
binding energy $E_{\rm bin} =E-M_0$ ($M_0$ is the total mass of the neutral
component)
in dependence on the form factor cutoff $\Lambda$.
In addition we also indicate the probabilities $P$ for having the
neutral $|0\rangle$ or the charged $|\pm\rangle$ components either
in $S$-- or $D$--wave in the bound state wave function. The size of the
system is characterized by the rms radius of the dominant neutral
$S$--wave component.

As $\Lambda$ is growing the binding energy becomes larger.
This is expected since the attractive potentials get stronger when $\Lambda$
is growing. The ratio between the neutral and the charged states is
very sensitive to the explicit value of $\Lambda$. The isospin-breaking effect
mainly comes from the mass difference between the neutral and the charged
$D\bar{D}^*$ system, that is $M_\pm-M_0 \approx$ 8.1 MeV.
The corresponding wave functions at a binding energy of $-0.40$ MeV are
shown in Fig.5. The neutral states dominate and the $D$--wave components
are negligible. In Fig.6 we also present the corresponding probabilities near
$|E_{\rm bin}|\approx0$. When the binding energy
become larger, however, the effect of the mass difference between neutral
and charged states weakens and the isospin symmetry is almost restored.
For example,
at a value of $\Lambda$=2500 MeV the binding energy is $-555.34$ MeV,
which by its absolute value is
larger than 8.1 MeV. For this value of $\Lambda $ the probabilities
$P(0_S)$,$P(0_D)$,$P(\pm_S)$ and
$P(\pm_D)$ are 46.1, 4.2, 45.6 and 4.1, respectively.

In Table~\ref{table:iso} we indicate our results in the isospin limit.
Compared with the full case of Table~\ref{table:isobreaking} the
isospin-breaking effects
reduce $\Lambda$ by about 90 MeV to obtain a bound state with the same
binding energy. The isovector
state of $D\bar{D}^*$ cannot make a bound state while the isoscalar state
can and mixing of isovector and isoscalar components leads to a weakening
of the binding energy.

To identify the relevant components both in the potential and the bound
$D \bar D^\ast$ configuration we also looked at reduced variants of this
approach. For example, when just keeping pion-exchange in the potential
a bound state can be formed for values of $\Lambda \geq  1700$ MeV.
This result is consistent with the findings in Ref.~\cite{Thomas:2008ja}.
Turning on in addition $\sigma$ meson exchange slightly increases the binding
energy with the effect depending on the explicit mass value $m_{\sigma}$.
We vary the value of $m_{\sigma}$ from 200 MeV to 600 MeV.
When we consider $\pi$ and $\rho$ meson exchange we only get a bound state
for values of $\Lambda \geq  1250$ MeV. Further additional meson exchange
components do not introduce a significant effect.

When we turn off the charged $D \bar D^\ast$ components we assume
that the X meson is no more isoscalar,
but a bound state of neutral $D^0D^{*0}$ components only.
Including in this case pion-exchange only leads to a minimal value of
$\Lambda = 4450$ MeV for the case of binding. Neglecting in addition the
$D$-wave effect a bound state can only be generated for
$\Lambda \geq  5900$ MeV.
These results fully agree with the ones of Ref.~\cite{Liu:2008fh}.
If we include all meson exchanges but still neglect the charged component
a bound state is obtained at $\Lambda = 2050$ MeV. Further neglect of
the $D$--wave results in a minimal value of $\Lambda = 2300$ MeV
to form a bound state.

Switching off the $D$-wave results in a
binding energy of $-0.23$ MeV at $\Lambda = 1250$ MeV.
This value of $-0.23$ MeV is smaller than the value of $-6.32$ MeV which
is obtained in the full calculation.
Thus we conclude that to form a bound state the condition of $I=0$ plays an
important role as well as $\pi$ and $\rho$ meson exchanges, but
$S-D$ wave mixing effect is less significant.

\section{$BB^*$ bound states}

Heavy quark symmetry and the nonrelativistic
approximation is more reliable for heavy-light systems containing
a $b$ instead of a $c$ quark.
>From this point of view it is worthwhile to study if in analogy
to the $DD^\ast$ system $B$ and $B^*$ mesons can also form molecular
bound states.
In Table~\ref{table:bb} we indicate possible $BB^*$ bound states in
dependence on $\Lambda$.
The mass values of $m_B$=5279.4 MeV and $m_{B^*}$=5325.0 MeV
are used as input.
The mass differences between the neutral and charged $B^{(*)}$ mesons are not
known experimentally at the moment. Therefore, isospin--breaking effects
only arise from the mass differences of the exchanged mesons such
as $\pi$ or $\rho$. However, these mass differences for the exchanged
mesons cause very slight deformation in the potential and its effect is
negligible. Presently it is the best option to assume isospin symmetry.

We calculate combinations of possible $B\bar{B}^*$ bound states assumed to be
an isoscalar or an isovector with charge parity even or odd. $B\bar{B}^*$
is likely to form
a bound state both with
$J^{\rm PC}=1^{++}$ and $1^{+-}$ in the isoscalar limit for reasonable
values of $\Lambda$.
In the case of $DD^*$ no bound state with $J^{PC}=1^{+-}$ is possible for
a suitable value of $\Lambda$. Compared to the $D\bar{D}^*$ case the
heavier mass of the $B\bar{B}^*$ states leads to a reduction of momentum
and hence of the angular momentum terms which act repulsive, and therefore
binding becomes easier.

As evident from the first part (related to $J^{PC}=1^{++}$)
of Table~\ref{table:bb}
binding energies even increase for reduced values of $\Lambda < 700$ MeV.
As a matter of fact the parameter $\Lambda$ was introduced to regularize
the attractive delta--function in the effective potential.
The potential becomes deeper as $\Lambda$ increases with the form factor
approaching 1 when $\Lambda$
goes to infinity. However, as displayed in Fig.7(b) the potential gets
deeper even when $\Lambda$ becomes smaller. This effect happens
when $\Lambda$ is smaller than the mass of the exchanged meson.
This might not have
any significant physical meaning at this low $\Lambda$ values, but
it is questionable if one can simply ignore this behavior of
a deepening potential at low $\Lambda$. Despite of this
$B\bar{B}^*$ bound states can be formed for a reasonable values of $\Lambda$,
around 1 GeV, which is larger than the mass of the exchanged mesons.

\section{Conclusion}

In this paper we explore the possibility to generate bound states with
$D$ and $D^*$ mesons in a potential model.
Using the HHChPT Lagrangian we construct the effective potential
including isospin symmetry breaking and also S--D wave mixing.
In this effective potential
the whole light pseudoscalar and vector mesons play the role of exchange
particles between $D$ and $D^*$ mesons. Because of the heavy mass of the
$D\bar{D}^*$ system a nonrelativistic approximation is meaningful.
We therefore solve a coupled channel Sch\"odinger equation with specific
potentials that are given by the effective Lagrangian.
Binding of the $D\bar{D}^*$ system is obtained for reasonable
values of $\Lambda \simeq 1.2$ GeV, increase of this cutoff value
will generate deeper binding. Comparing this result to the isospin
symmetry limit the $DD^*$ bound state where isospin symmetry is broken needs
a somewhat larger $\Lambda$ but still within a reasonable range of values.
By switching on and off various factors we also demonstrated that the relevant
ingredients for binding are the $I=0$ condition
and $\pi$, $\rho$ meson exchanges.
The S--D wave mixing effect is essentially negligible.

We extended our method to the $B\bar{B}^*$ system to see if
a molecular bound state is possible. Bound states of $B\bar{B}^*$ are
formed both with $J^{\rm PC}=1^{++}$ and $1^{+-}$ in the
isoscalar limit. But we find that it is harder to make a $B\bar{B}^*$ bound
state in the isovector limit just as for $D\bar{D}^*$. If the mass
differences between the neutral and charged $B^{(*)}$ were known,
then we could apply our method to see if isospin breaking effects
play an important role.
But presently we assume isospin symmetry in calculating the
binding energies of the $B\bar{B}^*$--system.

It is remarkable that there could be two different states in a molecular
$B\bar{B}^*$ picture: with $J^{\rm PC}=1^{++}$ and $1^{+-}$
(both with $I=0$). For example when $\Lambda=1000$ MeV
the binding energy of $1^{++}$ is $-21.13$ MeV
while that of $1^{+-}$ is $-0.34$ MeV.
Future possible detection of these states can give strong support to
the molecular approach in the heavy meson sector.

\begin{acknowledgments}

This work was supported by the DFG under Contract
No. FA67/31-2 and No. GRK683. This research is also part of the
European Community-Research Infrastructure Integrating Activity
``Study of Strongly Interacting Matter'' (HadronPhysics2,
Grant Agreement No. 227431) and of the President grant of Russia
``Scientific Schools''  No. 871.2008.2.
The work is partially supported by Russian Science and Innovations
Federal Agency under contract  No 02.740.11.0238.
We thank C.~Thomas and
V.~Ledoux for helpful discussions on the numerical calculations.

\end{acknowledgments}

\newpage

\begin{figure}\label{fig:diag1}
\centering{\
\epsfig{figure=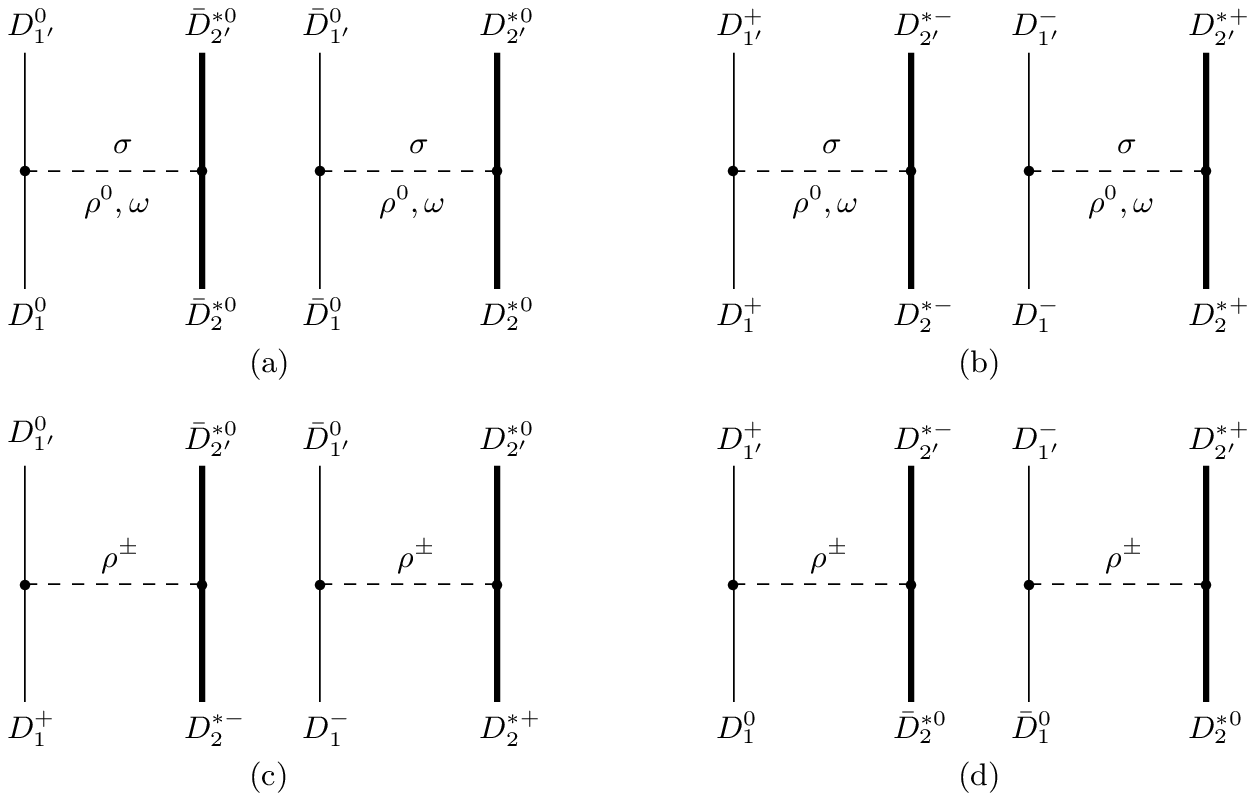,scale=1}}
\caption{
Feynman diagrams describing the scattering of $D$ and $D^\ast$
in the pseudoscalar-vector direct channel:
(a) and (b) -- Isospin channel $I=0$ and the isotopic factors
are equal to $-1$;
(c) and (d) -- Isospin channel $I=1$ and the isotopic
factors are equal to $2$.
The bold (solid) lines stand for vector $D^*$ (scalar $D$) mesons while
the dashed lines represent the exchanged mesons.}
\end{figure}

\begin{figure}\label{fig:diag2}
\centering{\
\epsfig{figure=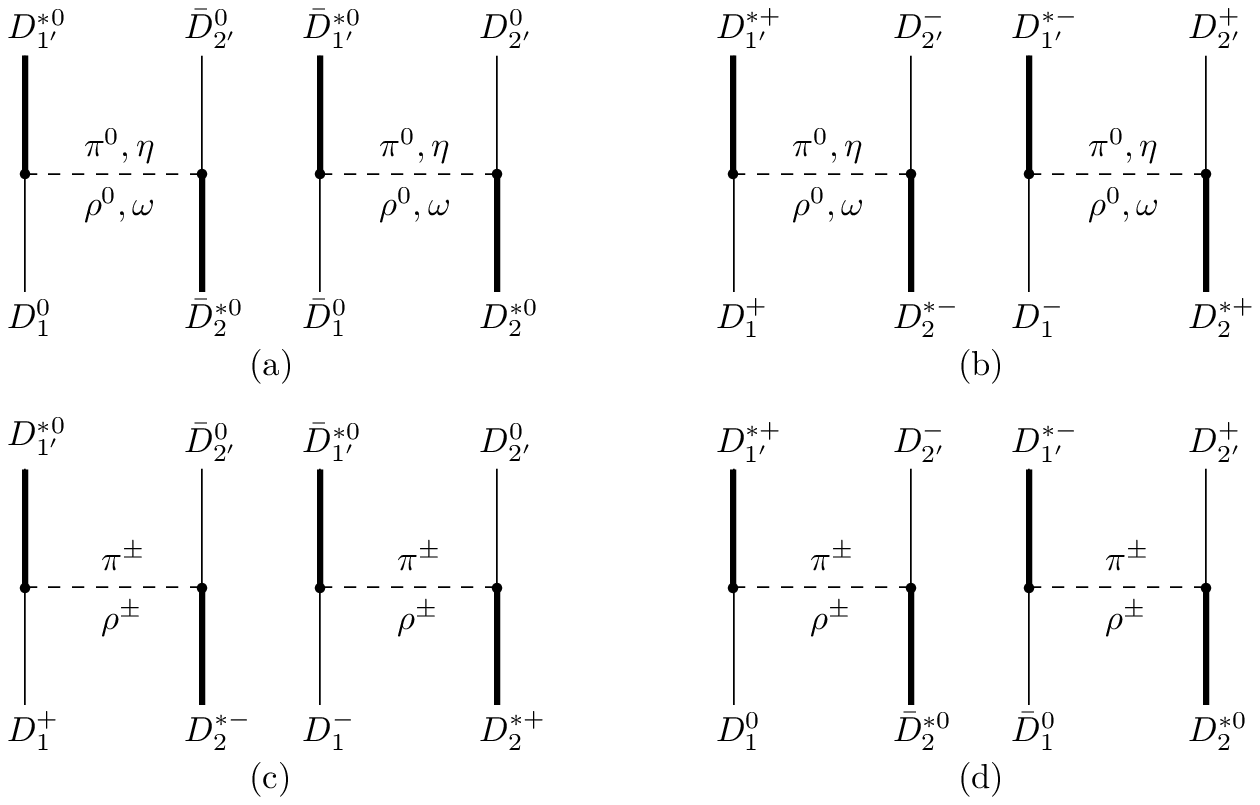,scale=1}}
\caption{Feynman diagrams describing the scattering of $D$ and $D^\ast$
in the pseudoscalar-vector crossed channel;  (a) and (b)
-- Isospin channel $I=0$ and the isotopic factors are equal to $-1$;
(c) and (d) -- Isospin channel $I=1$ and the isotopic factors
are equal to $2$.
The bold (solid) lines stand for vector (scalar) mesons
while the dashed lines represent the exchanged mesons.}
\end{figure}

\begin{figure}
\centering
\label{fig:PQCT}
\includegraphics[height=30em,scale=1.0]{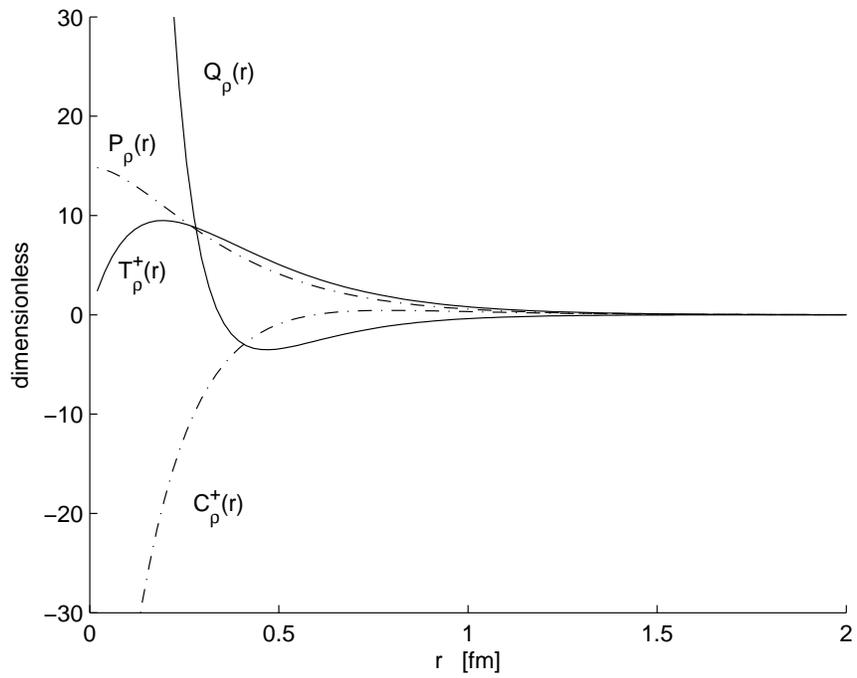}
\caption{Comparison of the shapes of the
dimensionless potentials for $\Lambda=1250$ MeV. The mass value is chosen as
the $\rho$ meson mass, $m=771$ MeV.}
\end{figure}

\begin{figure}
\label{fig:Cpotential}
\begin{tabular}{c}
\includegraphics[height=30em,scale=1.0]{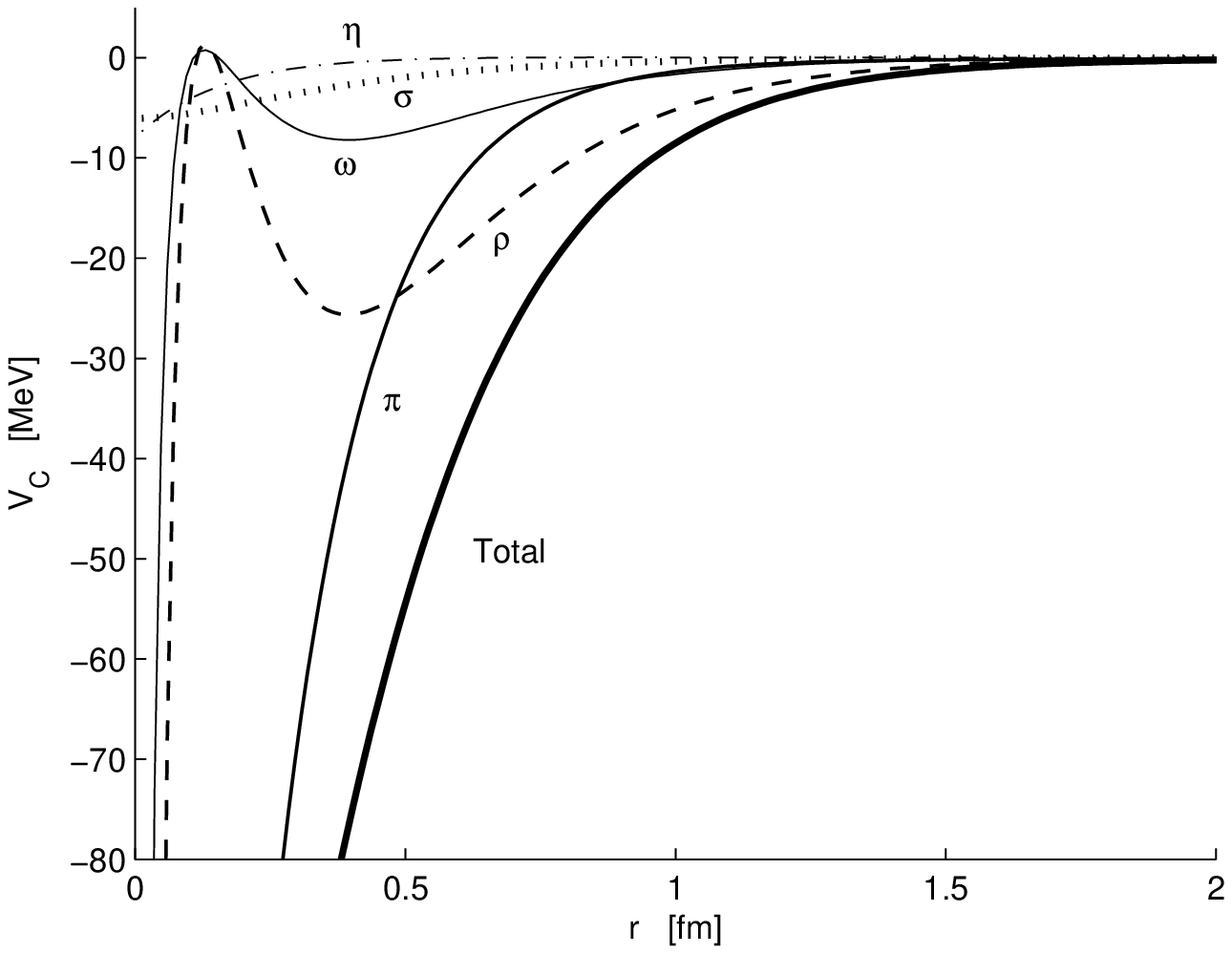}
\\(a)
\\
\includegraphics[height=30em,scale=1.0]{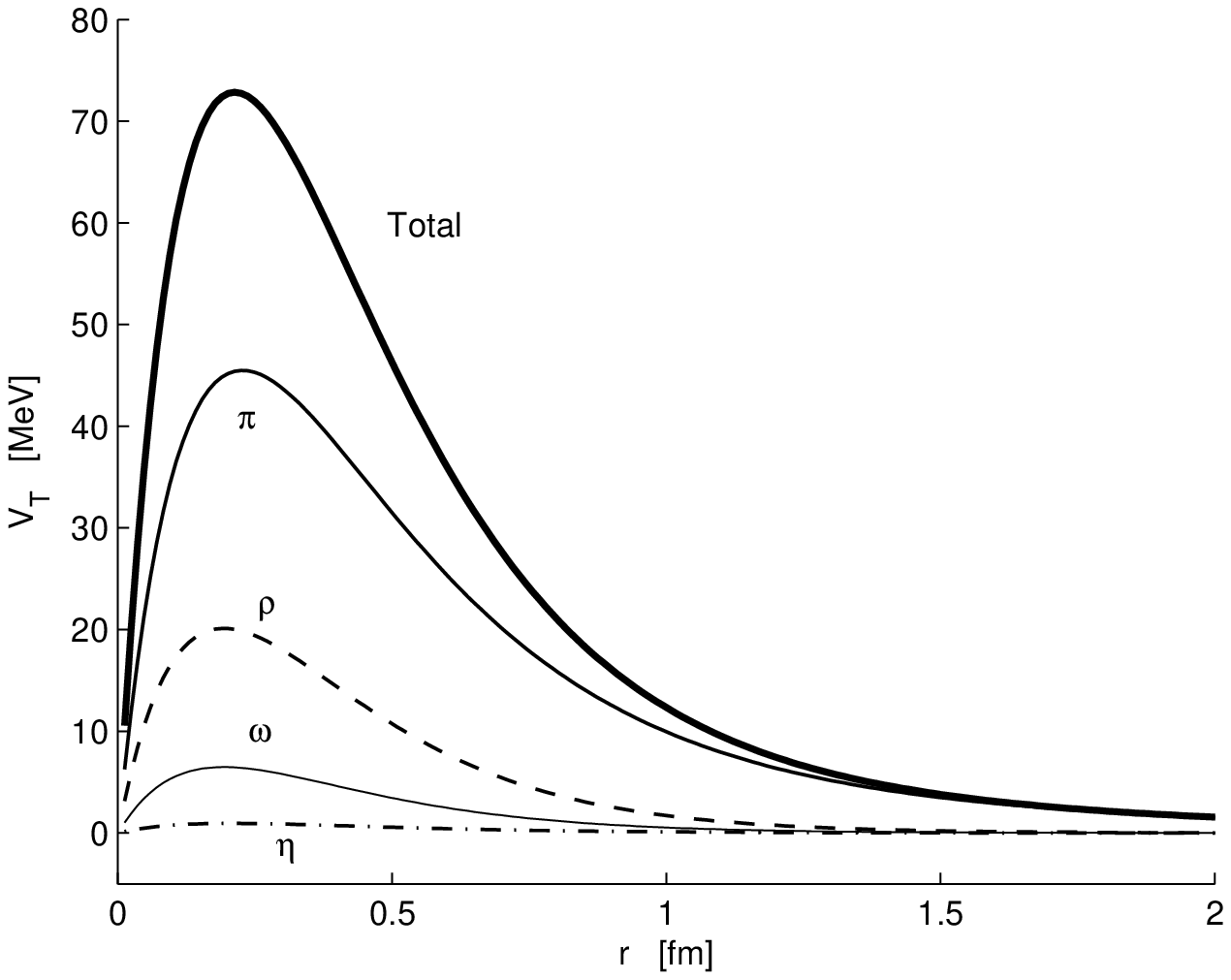}
\\
(b)
\end{tabular}
\caption{Central and tensor potentials due to meson--exchange
for $\Lambda$=1250 MeV.\\
(a) the central potentials $V_C$
($\pi$ and $\rho$ meson exchange is dominantly attractive);
(b) the tensor potentials $V_T$ due to the crossed
channels ($\pi$ and $\rho$ mesons are dominant and equal in sign).}
\end{figure}

\begin{figure}
\label{fig:wave_functions}
\centering
\includegraphics[height=30em,scale=1.0]{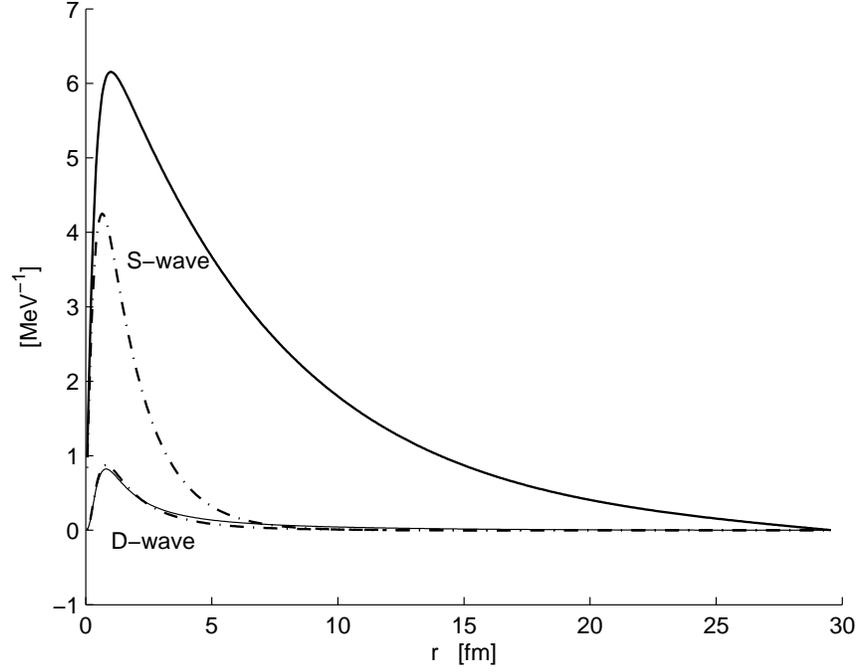}
\caption{Bound state wave function including isospin symmetry breaking
for $\Lambda$=1250 MeV. Bold lines denote the neutral states while
dot-dashed lines denote the charged states.}
\end{figure}

\begin{figure}
\label{fig:near_binding}
\centering
\includegraphics[height=30em,scale=1.0]{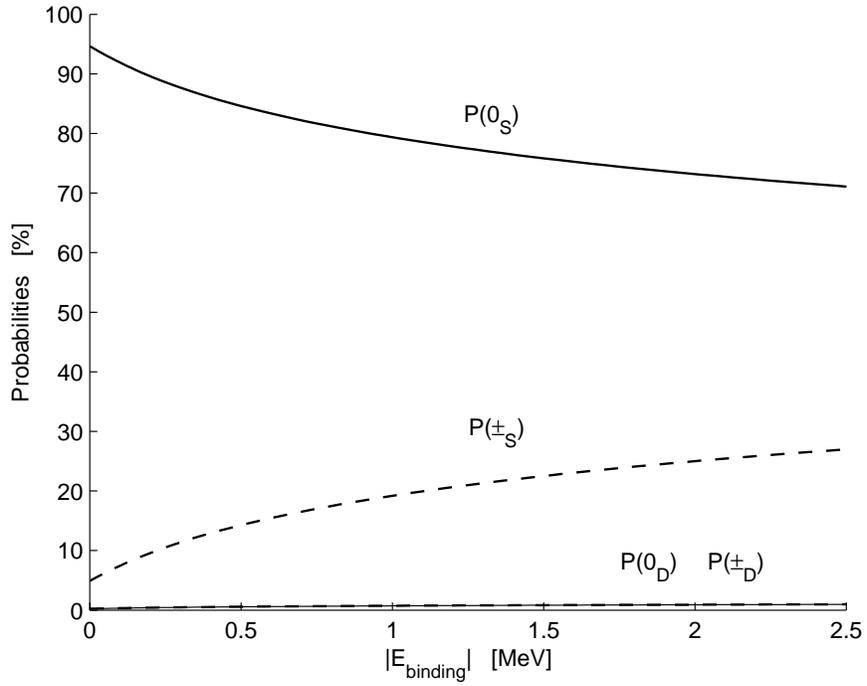}
\caption{The corresponding probabilities near $|E_{\rm bin}|\approx0$.
Bold lines denote the neutral states while dot-dashed lines denote
the charged states.}
\end{figure}

\begin{figure}
\label{fig:potential change}
\begin{tabular}{c}
\includegraphics[height=30em,scale=1.0]{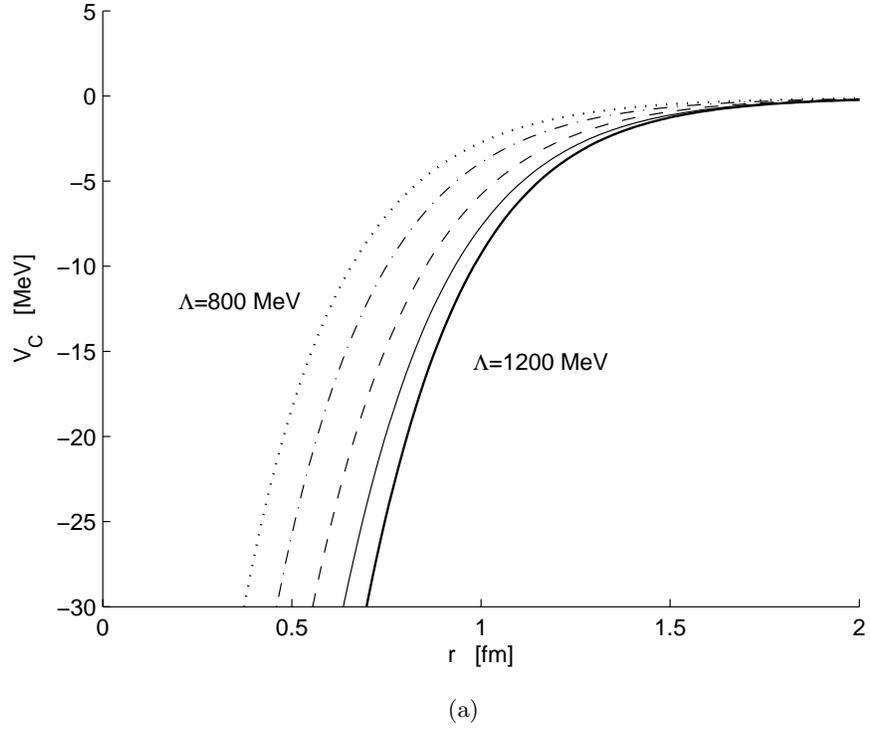}
\\
(a)
\\
\includegraphics[height=30em,scale=1.0]{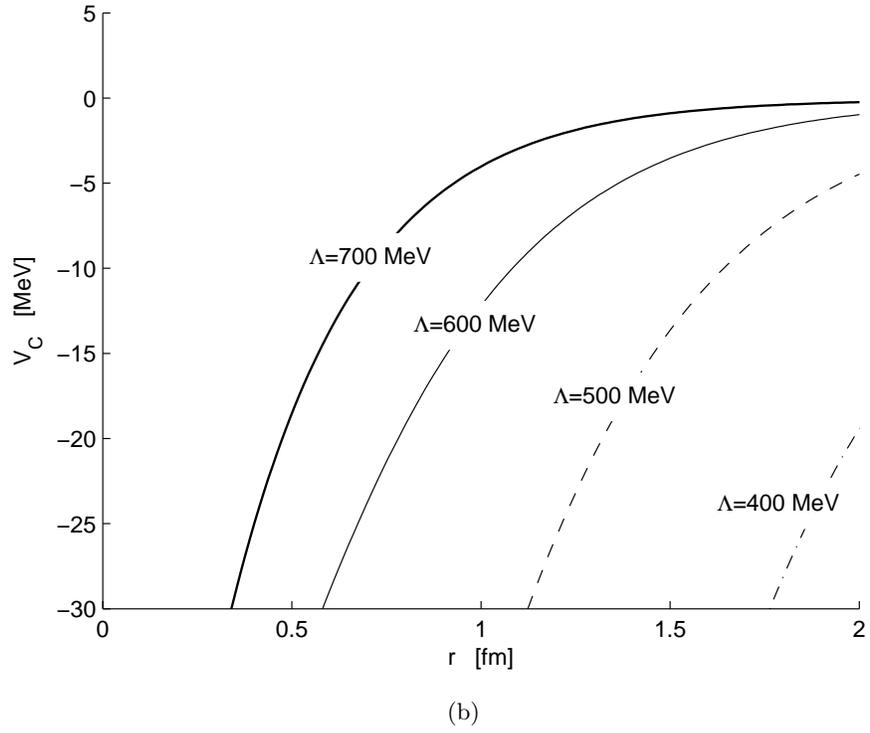}
\\
(b)
\end{tabular}
\caption{The behavior of the central potentials for various values of
$\Lambda$.}
\end{figure}

\newpage

\begin{table}
\caption{Binding energies with varying $\Lambda$ in the isospin
breaking case.}
\label{table:isobreaking}
\centering
\def\arraystretch{.8}
\begin{tabular}{c|c|c|c|c|c|c|c}
\hline
$\Lambda$ [MeV] & $E_{\rm bin}$ [MeV] & $E=M_0+E_{\rm bin}$ [MeV] & $P(0_S)
\%$&$P(0_D)\%$&$P(\pm_S)\%$&$P(\pm_D)\%$&rms$(0_S)$ [fm]\\
\hline
1100&No bound state&-&-&-&-&-&-\\
1136&$-0.10$&3871.10&91.9&0.3&7.5&0.3&7.5\\
1150&$-0.40$&3870.80&86.0&0.5&13.0&0.5&4.8\\
1160&$-0.71$&3870.49&82.1&0.6&16.7&0.7&3.7\\
1168&$-1.02$&3870.18&79.2&0.7&19.3&0.7&3.1\\
1200&$-2.65$&3868.55&70.5&1.0&27.5&1.0&1.9\\
1250&$-6.32$&3864.88&62.6&1.3&34.9&1.3&1.3\\
1300&$-11.10$&3860.10&58.2&1.5&38.8&1.5&1.0\\
1350&$-16.87$&3854.33&55.5&1.7&41.1&1.7&0.8\\
\hline
\end{tabular}

\vspace*{.2cm}

\caption{Binding energies with varying $\Lambda$ in the isospin
symmetry limit. }
\centering
\label{table:iso}
\def\arraystretch{.8}
\begin{tabular}{c|c|c|c|c|c|c|c}
\hline
$\Lambda$ [MeV] & $E_{\rm bin}$ [MeV] & $E=M_0+E_{\rm bin}$ [MeV] & $P(0_S)
\%$&$P(0_D)\%$&$P(\pm_S)\%$&$P(\pm_D)\%$&rms$(0_S)$ [fm]\\
\hline
1000&No bound state&-&-&-&-&-&-\\
1050&$-0.06$&3871.14&49.7&0.3&49.7&0.3&5.9\\
1076&$-0.41$&3870.79&49.5&0.5&49.5&0.5&3.7\\
1090&$-0.71$&3870.49&49.4&0.6&49.4&0.6&2.9\\
1100&$-0.97$&3870.23&49.3&0.7&49.3&0.7&2.5\\
1150&$-2.94$&3868.26&49.1&0.9&49.1&0.9&1.6\\
1200&$-5.95$&3865.25&48.9&1.1&48.9&1.1&1.2\\
1250&$-9.98$&3861.22&48.7&1.3&48.7&1.3&0.9\\
\hline
\end{tabular}

\vspace*{.2cm}

\caption{Binding energies of the $BB^*$ system.}
\label{table:bb}
\centering
\def\arraystretch{.8}
\begin{tabular}{c|c|c|c|c|c|c|c}
\hline
Isospin & $J^{\rm PC}$& $\Lambda$ [MeV] & $E_{\rm bin}$ [MeV]&$E=M_0+E_{\rm bin}$ [MeV]
        & $P(S) \%$&$P(D)\%$&rms$(S)$ [fm]\\
\hline
   &        &600&$-6.32$&10598.08&97.90&2.10&1.24\\
   &        &650&$-1.67$&10602.73&97.65&2.35&1.91\\
   &        &700&$-0.39$&10604.01&97.86&2.14&3.41\\
   &$1^{++}$&750&$-0.33$&10604.07&97.72&2.28&3.62\\
   &        &800&$-1.09$&10603.31&96.90&3.10&2.15\\
   &        &850&$-3.28$&10601.12&96.32&3.68&1.38\\
   &        &900&$-7.33$&10597.07&95.97&4.03&1.02\\
\cline{2-8}
I=0&        &950&No bound state&-&-&-&-\\
   &        &1000&$-0.34$&10604.06&92.70&7.30&3.92\\
   &        &1050&$-2.01$&10602.39&86.24&13.76&1.84\\
   &$1^{+-}$&1100&$-5.88$&10598.52&82.06&17.94&1.20\\
   &        &1150&$-12.72$&10591.68&79.45&20.55&0.89\\
   &        &1200&$-23.22$&10581.18&77.83&22.18&0.71\\
   &        &1250&$-38.00$&10566.40&76.80&23.20&0.59\\
\hline
   &        &4700&No bound state&-&-&-&-\\
   &        &4750&$-0.01$&10604.39&98.98&1.02&9.01\\
   &        &4800&$-0.06$&10604.34&98.40&1.60&7.00\\
   &$1^{++}$&4850&$-0.14$&10604.26&97.70&2.30&5.25\\
   &        &4900&$-0.26$&10604.14&96.97&3.03&4.03\\
   &        &4950&$-0.41$&10603.99&96.24&3.76&3.22\\
   &        &5000&$-0.62$&10603.78&95.52&4.48&2.67\\
\cline{2-8}
I=1&        &1950&No bound state&-&-&-&-\\
   &        &2000&$-0.04$&10604.36&99.45&0.55&7.47\\
   &        &2050&$-0.17$&10604.23&99.10&0.90&4.75\\
   &$1^{+-}$&2100&$-0.37$&10604.03&98.79&1.21&3.28\\
   &        &2150&$-0.67$&10603.73&98.53&1.47&2.50\\
   &        &2200&$-1.06$&10603.34&98.30&1.70&2.02\\
   &        &2250&$-1.55$&10602.85&98.11&1.89&1.70\\
\hline
\end{tabular}
\end{table}

\end{document}